\documentstyle[11pt,aaspp4,
flushrt]{article}

\slugcomment{To appear in The Astrophysical Journal}

\lefthead{Gonzalez et al.}
\righthead{The Velocity Function of Galaxies}

\def\gsim {\lower .1ex\hbox{\rlap{\raise .6ex\hbox{\hskip .3ex
        {\ifmmode{\scriptscriptstyle >}\else
                {$\scriptscriptstyle >$}\fi}}}
        \kern -.4ex{\ifmmode{\scriptscriptstyle \sim}\else
                {$\scriptscriptstyle\sim$}\fi}}}

\def\lsim {\lower .1ex\hbox{\rlap{\raise .6ex\hbox{\hskip .3ex
        {\ifmmode{\scriptscriptstyle <}\else
                {$\scriptscriptstyle <$}\fi}}}
        \kern -.4ex{\ifmmode{\scriptscriptstyle \sim}\else
                {$\scriptscriptstyle\sim$}\fi}}}
\def\cvir{c_{\tiny {\rm vir}}}
\def\Msun{M_{\odot}}
\def\hMsun{h^{-1}\Msun}
\def\hMpc{h^{-1}\rm{Mpc}}
\def\vch{v_c^{\rm halo}}

\begin{document}

\title{The Velocity Function of Galaxies}

\author{Anthony H. Gonzalez and Kurtis A. Williams} 
\affil{Department of Astronomy and Astrophysics}
\affil{ University of California at Santa Cruz}
\affil{ Santa Cruz, CA 95064}
\author{James S. Bullock, Tsafrir S. Kolatt, and Joel R. Primack} 
\affil{Department of Physics}
\affil{ University of California at Santa Cruz}
\affil{ Santa Cruz, CA 95064}


\begin{abstract}
 We present a galaxy circular velocity function, $\Psi(\log v)$,
derived from existing luminosity functions and luminosity-velocity
relations.  Such a velocity function is desirable for several
reasons. First, it enables an objective comparison of luminosity
functions obtained in different bands and for different galaxy
morphologies, with a statistical correction for dust extinction. In
addition, the velocity function simplifies comparison of observations
with predictions from high-resolution cosmological N-body simulations.

We derive velocity functions from five different data sets and find
rough agreement among them, but about a factor of 2 variation in
amplitude. These velocity functions are then compared with N-body
simulations of a $\Lambda$CDM model (corrected for baryonic infall) in
order to demonstrate both the utility and current limitations of this
approach. The number density of dark matter halos and the slope of the
velocity function near $v_*$, the circular velocity corresponding to
an $\sim L_*$ spiral galaxy, are found to be comparable to that of
observed galaxies. The primary sources of uncertainty in construction
of $\Psi(\log v)$ from observations and N-body simulations are
discussed and explanations are suggected to account for these
discrepancies.

\end{abstract}


\section{Introduction}

 Galaxy luminosity functions and the Tully-Fisher (TF) relation are
key tools for testing models of galaxy formation and incorporating
them into a larger picture of gravitational structure formation. An ultimate
goal is to be able to reproduce these
quantities starting from cosmological N-body simulations.
A significant complication is that observed galaxy luminosities
are dependent upon a number of astrophysical processes (e.g., star
formation history, gas cooling, internal extinction, supernovae feedback,
chemical evolution, gas reheating and sharing between galaxies, stellar
mass functions, etc.). These factors, which are generally poorly constrained,
obscure the connection between formation processes and observable quantities.

  These complications have not however deterred attempts to bridge the
gap between the dark matter halos generated by N-body simulations and
observed galaxies. In the past decade, the use of semi-analytic models (SAMs), 
which create galaxies from dark matter halos by modelling the relevant baryonic
physics as global galaxy properties, has become the favored technique for
tackling this issue. SAMs have had impressive success in reproducing
both observed luminosity functions and TF relations, although not always both
at the same time (\cite{kau93,col94,som99}). A limitation
of this approach is that the current models necessarily contain many
degrees of freedom, and a number of aspects of the models are 
oversimplified (\cite{som99}).

An alternative approach that complements the SAMs is the use of observational
data to generate quantities that may be linked more directly with dissipationless N-body
simulations.
One such quantity is the galaxy velocity function, 
$\Psi(\log v_c)$, which describes the number density of galaxies per unit circular 
velocity. $\Psi(\log v_c)$ can be constructed using published luminosity 
functions and luminosity-velocity ($l-v$) relations.
The velocity function is valuable for several reasons. First, conversion of
luminosity functions into velocity functions places surveys obtained in different
bands on equal footing (with the caveats discussed in \S4.2). This permits direct comparison of
the surveys and provides a single target function for which the simulations can aim.
Second, by removing the need to model luminosity or understand the physical origin of
the TF relation, a number of processes modelled by standard SAMs can be ignored.
Only processes that affect baryonic infall,
and hence the gravitational potential, impact the velocity function (see \S4.3). These include
gas cooling and supernovae feedback.  For these reasons, the velocity function can be a
useful tool for probing the connection between large scale gravitational physics and 
galaxy formation when coupled with the latest generation of cosmological N-body simulations.

Construction of a velocity function was suggested by Cole and Kaiser (1989), and an
empirical velocity function was created by Shimasaku (1993). The latter work utilized a sample
of nearby, bright galaxies from the {\it{Third Reference Catalogue of Bright Galaxies}} 
(de Vaucouleurs et al. 1991),  with velocities derived from a combination of 21 cm observations
and $l$-$v$ relations. Interestingly, Shimasaku also extended this analysis to attempt to include
clusters, finding that the galaxy and cluster velocity functions are consistent with being
derived from a single dynamical population.

The goal of this paper is to determine $\Psi(\log v_c)$.  The approach
taken will be first to use the luminosity function from a single 
survey (SSRS2) and a
single set of $l-v$ relations to create a detailed velocity
function. This analysis will be used to assess the importance of
correctly accounting for factors that may alter the resultant
$\Psi(\log v_c)$, such as internal galactic extinction.  The results will
then be used as a foundation for production of simplified velocity
functions for a variety of surveys and $l-v$ relations to assess the
robustness of the results. It will be demonstrated that the derived
velocity function is 
robust to within a factor of 2 for  $70\lsim v_c\lsim 260$ km s$^{-1}$.
To illustrate this we use the results from
Adaptive Refinement Tree simulations (ART, \cite{kra97,kly99}).
Comparison with these results
shows agreement between observations and theory at $v\sim v_*$, but an
excess of dark matter halos at lower velocities. 

With the goal of fostering improvement beyond the current work, 
we also attempt to identify here the primary observational 
and theoretical sources of uncertainty. Among the observational 
limitations, uncertainties associated with the spiral and 
elliptical $l-v$ relations prove to be the most significant 
factors. Among the theoretical issues, correction of 
dissipationless models for the uncertain effects of baryonic 
infall is one of the most significant sources of uncertainty. 

In \S\ref{sec:schechter} we define the velocity function. The various
ingredients necessary to construct this function are detailed in
\S\ref{sec:ingredients}. Then we
turn to the data analysis in \S\ref{sec:analysis}, where we also compare
the derived velocity functions with the N-body simulation and discuss 
sources of uncertainty. We discuss our
results and conclusions in \S\ref{sec:conc}.
Throughout this paper the Hubble parameter is expressed
as $H_0$=100 $h$ km s$^{-1}$ Mpc$^{-1}$.

\section{Schechter Velocity Functions}
\label{sec:schechter}

Galaxy luminosity functions are normally parameterized using a 
Schechter function (\cite{sch76}) of the form

\begin{equation}
 \Phi(L) dL = \Phi_* \left(\frac{L}{L_*}\right)^\alpha
 \exp{\left(-\frac{L}{L_*}\right)} \frac{dL}{L_*},
\end{equation}
where the three observationally determined parameters $\Phi_*$,
$\alpha$, and $L_*$ respectively describe the normalization, faint end
slope, and break point of the luminosity function. If the velocity, $v$, is
related to $L$ by a simple power law ($L=A v^n$), the number
density of galaxies per unit velocity can also be
described by a generalized form of the Schechter function,

\begin{equation}
 \tilde{\Psi}(v) dv = \tilde{\Psi}_* \left(\frac{v}{v_*}\right)^{\beta}
 \exp{\left[-\left(\frac{v}{v_*}\right)^n\right]} \frac{dv}{v_*},
\end{equation}
where $v_*=(L_*/A)^{1/n}$, $\beta\equiv n (\alpha+1)-1$, and $\tilde{\Psi}_*\equiv n \Phi_*$ 
(see Appendix).
Equivalently, this can be expressed in terms of $\eta=\log(v)$ as,

\begin{equation}
 \Psi(\eta) d\eta = \Psi_* \left(10^{\eta-\eta_*}\right)^{\beta+1}
\exp{\left[-\left(10^{\eta-\eta_*}\right)^n\right]} d\eta,
\end{equation}
where $\Psi_* = (\ln10)\tilde{\Psi}_*$. 
Luminosity-velocity relations exhibit tight
correlation, so we choose to construct a circular velocity
function for comparison with cosmological models. Specifically, we
define $v_c$ to be the circular velocity measured in the flat part of
a spiral galaxy's rotation curve. For spirals $v_c$ can be observed directly;
for ellipticals the velocity dispersion, $\sigma$, is
observed, and so it is necessary to convert $\sigma$ to $v_c$.
In this paper, the simplyfing assumption is made
that an elliptical can be modeled as an isothermal sphere, in which
case $v_c=\sqrt{2} \sigma$ (\cite{bt}). The quantity $v_c$ can also be
extracted from 
very high-resolution N-body simulations, so a direct comparison 
of observed and simulated $\Psi(\log v_c)$ is possible.

\section{Velocity Function Ingredients}
\label{sec:ingredients}
\subsection{Survey Luminosity Functions}
     
We impose several criteria on the input luminosity function to
simplify the analysis. First, it is preferable that 
the selected survey contain a large number of galaxies, 
encompass a large volume, and extend to luminosities well below
$L_*$.  Second, morphological
information is necessary since spirals, ellipticals, and irregulars
are observed to follow different $l-v$ relations.
Table 1 lists a number of recent
surveys for which luminosity functions have been computed, as well as Schecter
parameters and the approximate magnitude range over which the Schechter fit is
valid.  Only the CfA2, SSRS2, and APM surveys meet the above criteria
(\cite{mar94,mar98}). The APM luminosity functions derived for different morphological
types have been called into question by several groups however (\cite{mar94,zuc94}), 
and so we refrain from using this morphological data.  Also, although it lacks morphological
information, the Las Campanas Redshift Survey (LCRS) spectroscopically
subdivides the galaxy population in a manner that roughly corresponds
to morphological types (\cite{bro98}).

\placetable{tbl-1}

There are other practical considerations regarding the $l-v$
relations that impose further constraints on which surveys can be
utilized. In particular, the CfA survey uses the Zwicky magnitude
system, for which $l-v$ relations are not published for any
morphological type.
The situation is slightly better for the $R$-band
LCRS; the $R$-band Tully-Fisher relation is well-studied and exhibits a
tight correlation, but similar information is not available for
ellipticals or irregulars.  In fact, $l-v$ relations have thus far
been published for all morphological types only in the
$B$-band. Consequently, for our initial effort at generating $\Psi(\log v_c)$
we select the $B$-band SSRS2 survey.  We use the Schechter
luminosity function parameters derived for the SSRS2 survey assuming 
no Virgocentric
infall, and note that Virgocentric infall corrections to the luminosity function
have only a modest effect on the results (\cite{mar98}).

\subsection{Luminosity-Velocity relations}

The observational limitations of luminosity-velocity relations pose the
greatest challenge to construction of the velocity
function. Derivation of $\Psi(\log v_c)$ requires that well-calibrated
$l-v$ relations exist for all morphological types that contribute
significantly to the luminosity function.

For spirals the forward Tully-Fisher (TF) relation 
($M = a-b[\log 2v_c -2.5]$)
has been extensively
studied (\cite{tul77}). Independent analyses have generated 
consistent results in the
$I$- and $R$-bands and have demonstrated that the intrinsic scatter in the
relation is $\lsim$0.4 mag at these wavelengths (\cite{wil96}).  In
the $B$-band less effort has been expended towards calibration of the TF
relation because the observed scatter is greater than at longer
wavelengths.  Still, several calibrations have been published. In
particular, the TF relation derived in the work of Yasuda, Fukugita, and
Okamura (1997) is chosen for construction of the SSRS2 velocity function.
Table 2 lists TF parameters derived by various authors in $B$, $R$, and $K$.
The different $B$-band relations will be used to evaluate the effect of
the choice of TF parameters on the derived $\Psi(\log v_c)$.
Also given in Table 2 is the velocity range spanned by the data used to define
each relation. It is important to note that in no case has the TF relation
been defined above $\sim$350 km s$^{-1}$. Data is also sparse below $\sim$100
km s$^{-1}$, but the work that has been done for both spirals and dwarfs at
lower circular velocities indicates that there is no dramatic departure from the
TF relation down to $v_c\lsim$20 km s$^{-1}$  (\cite{hof96,ric87}).

\placetable{tbl-2}

For ellipticals the $D_n$-$\sigma$ relation is the most accurate means
of converting luminosity to velocity dispersion
(\cite{dre87}). However, SSRS2 and other existing large surveys do not
publish effective radii and luminosities for individual
galaxies. Since the SSRS2 gives only a luminosity function for the
E/S0 population, the Faber-Jackson 
(FJ, \cite{fab76})
relation must be
employed. Unfortunately, subsequent to the development of the
$D_n$-$\sigma$ relation scant effort has been given to calibration of
the FJ relation, and so we refer to early work by 
\cite{dev82}.  Using a sample of 86 E's and S0's
with recessional velocities greater than 1550 km s$^{-1}$ and 135$<\sigma<$376
km s$^{-1}$, these authors
observe that

\begin{equation}
L_{B_T}\propto \sigma^{3.08\pm0.28},
\end{equation}

or 

\begin{equation}
M_{B_T}= (-19.71\pm0.08) +(7.7\pm0.7) \left(\log \sigma -2.3\right) + 5 \log h .\hspace{1cm}  ``Best"
\end{equation}
Although this is the relation used for the SSRS2 analysis, we caution
that the statistical error may be a significant underestimate of the
uncertainty in this relation. A larger sample of pure E's in the same
paper yields 

\begin{equation}
M_{B_T}= (-19.38\pm0.08) +(9.0\pm0.7) \left(\log \sigma -2.3\right) + 5 \log h .
\hspace{1cm}  ``High"
\end{equation}
For the remainder of the paper these will respectively be denoted as
the ``Best'' and ``High'' FJ relations, as they represent
our best estimate of the true relation and the relation with the
highest probable slope.  The error associated with the slope of the FJ
relation has a negligible effect on the results, but the difference in 
zeropoints is a significant source of uncertainty.

For irregulars, the $l-v$ relationship remains poorly
constrained. Fortunately, this does not impede the 
calculation of $\Psi(\log v_c)$
because irregulars are only a trace population in the velocity regime
($v_c \gsim 100$ km s$^{-1}$) 
probed by current cosmological simulations. For completeness, we transform the
irregular population using the $l-v$ relation  recently derived from a sample
of 70 dwarf irregulars (\cite{hof96}). This relation has the same form
as the spiral TF relation, as well as a similar slope 
($6.50\pm0.63$ for the irregulars as compared to $6.76\pm0.63$ for the
TF relation of \cite{yas97}).

\subsection{Passband Effects} 
A complication in combining the published TF and FJ relations with survey 
luminosity functions is that slightly different filters have been
used by different authors. For example, the Yasuda et al. (1997) TF relation was 
obtained in $B_T$, while the APM and UKST surveys use $b_J$. Table 3 gives the
passband transformations (and references) used to transform the data onto 
comparable photometric systems. Of particular note is the $R$-band data. The LCRS survey
data was obtained in $r_g$, but calibrated to the $R_C$ system. Further, the TF data
of Courteau was obtained using a Spinrad $r$ filter, but calibrated to the $r_g$ system.
The color transformation given in Table 3 to systhesize these two data sets is derived 
from the transformations given by Djorgovski (1985), utilizing the galaxy color information 
of Fukugita, Shimasaku, \& Ichikawa (1997).
Correction for the effect of different passbands has a noticeable effect upon the 
results (e.g., $v_*$ increases by $\sim$ 20 km s$^{-1}$ for SSRS2), and so must not be 
ignored. 
\placetable{tbl-3}

\subsection{Internal Extinction} 
A bias that is normally ignored in creation of luminosity functions is
dimming due to internal absorption.  While absorption is presumably
negligible in ellipticals, internal absorption is significant in
spirals and is a function of both inclination and luminosity
(\cite{gio95,tul98}). Luminosity functions are inherently averaged
over inclination and contain galaxies spanning a wide range of
luminosity, whereas TF relations are generally calibrated with bright
spirals for which the magnitudes are corrected to face-on.
Consequently, it is necessary to correct the luminosity function for
internal absorption before using the TF relation to transform
luminosity to velocity. 

The issue of internal extinction in spirals has been a source of significant debate 
since Valentijn's (1990) claim that spiral galaxies are optically thick. This work
was later challenged by Burstein et al. (1991) and Davies et al. (1993), 
who argued that Valentijn's study was subject to significant selection effects and 
biases.{\footnote{For a nice discussion of this topic, see Bottinelli et al. (1995).}}
Subsequently, there has been mounting evidence, if not a consensus, that
extinction is significant, but not as extreme as suggested by Valentijn
(\cite{gio95,bot95,cou96,tul98}). For this work, we adopt the extinction 
corrections of Tully et al. (1998, hereafter T98). Our motivation is threefold. First,
and most importantly from a practical perspective, only T98 have published 
extinction corrections in $B$, $R$, and $K$ --- the three bands for which luminosity functions
are available. Second, this is the only work other than Giovanelli et al. (1995, 
hereafter G95) that models dependence of absorption on luminosity. {\footnote{Such
a dependence is expected, and is closely linked to the morphological type dependence given in the
Third Reference Catalogue (\cite{dev91}).}} Third, unlike Botinelli et al. (1995, 
hereafter B95) and G95, who concentrate on late types (Sc-Sd), the Tully sample 
probes a broader range of spirals. In the $I$-band, a comparison of T98 and G95's 
results shows consistent trends, but greater extinction in G95, possibly due to 
the sample composition (\cite{tul98}). Similarly, B95's average extinction is 
higher than for T98. A comparison of the effects of using B95 and T98 is shown in Figure 1(a) 
(see \S\ref{sec:analysis}).
While the correspondence is good, we caution that this topic remains far from settled, 
and the extinction correction is one of the most significant sources of uncertainty
in constructing $\Psi(\log v)$.

When averaged over inclination, assuming randomly distributed
inclination angles,{\footnote {We assume, as in Tully et al. (1998), 
that the aspect ratio of an edge-on spiral is $b/a=0.2$}} 
the luminosity corrections derived from T98 are:

\begin{equation}
M_{B_J}^{cor}=\frac{1}{0.92} \left[ M_{B_J} + 0.08 (15.6+5\log h_{80} )\right], \hspace{1cm}M_{B_J}<-15.6
\end{equation}

\begin{equation}
M^{cor}_{R_C}=\frac{1}{0.95} \left[ M_{R_C} + 0.05 (16.2+5\log h_{80} )\right], \hspace{1cm}M_{R_C}<-16.2
\end{equation}

\begin{equation}
M^{cor}_{K'}=\frac{1}{0.99} \left[ M_{K'} + 0.01 (18.3+5\log h_{80} )\right], \hspace{1cm}M_{K'}<-18.3
\end{equation}

The formalism for inclusion of this luminosity dependent extinction
correction within the generalized Schechter function is given in the
Appendix.{\footnote{This is in fact a simplification. The observed
luminosity function is a convolution of the face-on luminosity
function with an extinction-induced, asymmetric magnitude dispersion
function. To account for extinction correctly, it is necessary to
deconvolve the luminosity function to face-on. This would slightly
flatten the faint end slope, but the effect is significantly smaller than
other sources of error, 
so we simply note the existence of
this generally overlooked effect. Also, we note that intrinsic scatter in
the TF and FJ relations lead to a similar effect. }}

\section{Analysis and Results}
\label{sec:analysis}
\subsection{SSRS2}
	With all the ingredients assembled, we now construct a
velocity function from the SSRS2 luminosity function. A first test is
to assess the impact of internal extinction in the spiral
population. Figure 1(a) shows the extinction-corrected and uncorrected
spiral velocity functions, $\Psi(\log v_c)$. To illustrate the impact of
the choice of extinction correction, both the T98 and B95 extinction laws
are applied. The net effect of both corrections is to shift
the function to higher velocity by approximately 30 km s$^{-1}$. The Tully 
correction is utilized in all subsequent figures.
Next, the impact of different choices for
TF and FJ relations is assessed. Figure 1(b)
shows that the relations published by Yasuda et al. (1997) and Richter
et al. (1987) are consistent, whereas the
disparate values for the FJ relation lead to a significant change at
high velocities. For the ``Best" relation, spirals provide the
greatest contribution at all velocities; for the ``High" relation,
ellipticals dominate above $v_*$. This is driven by the change in
zeropoint.

\placefigure{fig1}

	The total velocity function can be seen as the heavy solid curve
in Figure 1(c), with the two light solid lines tracing the central
curve indicating the uncertainty due to the formal ($1\sigma$) statistical errors from the
luminosity function and TF parameters. Also displayed are the
constituent velocity functions for each morphological type, using the Yasuda, Fukugita,
and Okamura (1997), deVaucouleur and Olson (1982), and Hoffman et al. (1996) $l-v$
relations for spirals, ellipticals/S0's, and irregulars, respectively.
Readily apparent is the dominance of the spiral population. Only at velocities
well above $v_*$ does the elliptical population contribute
substantially. Given this dominance, it is of interest to ask how the
total velocity function would differ under the assumption that all
galaxies are spirals.
Namely, how important is the segregation between morphological types
in the translation of luminosity to velocity?
From Figure 1(d) it can be seen that
this ``spiral approximation'' is quite good, altering the total
velocity function (Composite \#1) by $\sim 10\%$, less than the formal errors. The
validity of such an approximation is important if we wish to compare with
LF surveys that lack morphological information.

There are several important notes of caution which should be mentioned.
If the zeropoints of the $l-v$ relations are significantly in error, then at 
high velocities the elliptical population may dominate. To assess the 
magnitude of this effect, we plot Composite \#2 in Figure 1(d), which uses 
the ``High'' FJ relation. Composite \#2 has an $\sim$20\% higher amplitude than the 
spiral approximation near $v_*$, and has a steeper slope above $v_*$. 
Also, demonstration that the spiral population is 
dominant in the $B$-band SSRS2 does not assure that the same is true
for galaxy samples selected in other bands, as we may be observing 
substantially different galaxies (\cite{lov98}). For $R$-band and bluer bands, this
should be a mild effect. In comparison to $B$-band, ellipticals in the $R$-band are
$\sim$0.15 mag brighter relative to spirals (\cite{fuk97}). By
$K$-band however, the spiral approximation should be very poor. As
compared to $B$-band, ellipticals in $K$ are $\gsim$ 1 mag brighter relative to
spirals. Consequently, use of the spiral approximation will 
artificially inflate both the derived $v_*$ and $\Psi_*$. In the
next section we compare velocity functions derived from different
surveys in order to provide a lower limit for the systematic errors
that are no doubt present. For completeness we include a $K$-band
survey, which illustrates the breakdown of the spiral approximation.

\subsection{Survey Comparison}
	SSRS2 is the only survey for which it is possible to generate
velocity functions for each morphological type, and so it is necessary
to employ the spiral approximation if we wish to compare velocity
functions from different surveys. This is done for the SSRS2, APM,
UKST/Durham, and LCRS (\cite{mar98,lov92,rat98,bro98}), and also for a 
$K$-band survey by \cite{gar97}.
For the $B$-band surveys, the TF relation of Yasuda, Fukugita, \& Okamura (1997)
is utilized. For the $R$- and $K$-bands we use the work of Courteau (1997)
and Malhotra et al. (1996), respectively.  The values of the
parameters in each of these relations are given in Table 2.  The
resulting $\Psi(\log v_c)$
are displayed in Figure 2; Figure 3 shows the same data, but only in the regime
where the TF relation is also constrained. The
generalized Schechter parameters (Eq. 2) corresponding to these velocity
functions are given in Table 4.  \placetable{tbl-4} \placefigure{fig2}
\placefigure{fig3}

The $R$- and $B$-band data all agree within the quoted observational
errors. This can be seen by comparing the parameter values in Table
4. We also illustrate this in Figure 4 by plotting $\Psi_{240} = \Psi(\log v_c)$ at
$v_c = 240$ km s$^{-1}$
for each survey. One important note is that, for the
LCRS catalog, Figure 2 is more indicative of the actual agreement in
$\beta$ than is the value in Table 4. This is because the fit in
Figure 2 is the sum of Schechter functions used to fit individual
spectroscopic clans, whereas the value in Table 4 is derived using a
single Schechter fit which is a visibly bad match to the data at the
faint end (\cite{lin96,bro98}). We also note that a rough conversion 
of the CfA survey data to the $B$-band is possible via the
transformation $M_{B_T}=M_Z-0.45$ (\cite{sha84}). We do not plot the CfA velocity
function in Figure 2, but find that it is consistent with the $B$- and $R$-band velocity functions.
\placefigure{fig4}

The $K$-band data primarily serves to emphasize the limitations of this approach.
For $K$, both $v_*$ and $\Psi_*$ are high relative to the other
surveys. The quoted errors are large, but the difference in the
resulting velocity function is statistically significant. This is not 
surprising, as there are several reasons to expect the $K$-band velocity 
function to be discrepant.  First, and most importantly, the spiral 
approximation {\it {should}} break down in $K$, as discussed at the end of
\S4.1. Consequently, for future $K$-band
surveys morphological information will be necessary if they are to be
used to generate a velocity function.  In addition, 
the Gardner et al. (1998) $K$-band luminosity function assumes a value of
$q_0=0.02$. Recent supernovae surveys indicate that a more likely
value is $q_0\approx-0.5$ for a flat universe (\cite{per98}), and so
there is an additional uncertainty that we have not included in
the error budget, which may decrease $\Psi_*$ and increase $v_*$ by
$\sim$20\%.  

Furthermore, of the three bands used, the TF
relation is least well-established in the $K$-band.  Three notable
recent determinations of the $K$-band TF relation are provided by 
Malhotra et al. (1996), Tully et al. (1998), and de Grijs \& Peletier (1999).
The parameters for each are given in Table 2. Consistent slopes are found by
all three groups, but the zeropoint variation is large. The zeropoint from T98 is
dependent upon the assumed distance to Ursa Major (\cite{ver97}), and also no errors
are quoted. de Grijs \& Peletier quote a 1-$\sigma$ error of 1.58 magnitudes. 
Malhotra et al., whose TF relation we employ, have the smallest quoted errors, but their
TF relation is based on DIRBE observations of only 7 Local Group galaxies including the Milky Way.

\subsection{Comparison with Simulations}
Although the comparison of the velocity function with 
a halo velocity function derived from N-body 
simulations is more straightforward than a corresponding
luminosity function comparison, 
there are a few caveats.  The first of these concerns how
to assign an appropriate value of $v_c$ to each simulated
dark matter halo, and
the second concerns the association of 
very high- and low-velocity halos with galaxies.  

It has been known for some time that the density profiles
of simulated dark matter halos are not well-approximated by isothermal
spheres (see e.g., 
Navarro, Frenk, \& White 1996 and references therein).
Unlike isothermal spheres, which have flat circular
velocity curves, the maximum rotation velocities of halos are not
the same as their virial velocities.
Galaxy formation also affects dark matter halo velocity curves 
due to the infall of cool baryons.
There are thus at least three possibilities for the $v_c$ to use in constructing
a halo velocity function:
\begin{enumerate}
\item $v_c = v_{\rm vir} \equiv \sqrt{G M_{\rm vir}/ R_{\rm vir}}$, the
      circular velocity of the halo at its virial radius, $R_{\rm vir}$.
\item $v_c = v_{\rm max}$, the maximum rotation velocity of the halo. 
\item $v_c = v_{\rm max}^{\rm corr}$, the maximum velocity of the halo 
      after baryonic infall has occurred.
\end{enumerate}
Clearly, option 1 is inappropriate. 
Recall that $v_c$ is the circular velocity measured at the flat part of a
disk galaxy rotation curve.
Halo velocity curves typically
flatten at $10-20\%$ of the virial radius, and $v_{\rm vir}$ may be 
as small as $60\%$ of $v_{\rm max}$ (\cite{bul99a}).  
So, although halo $v_{\rm vir}$ 
velocity functions are the most straightforward of the three
to estimate, we will
not do so here in order to focus on the more appropriate options.
Choice 2 is  more sensible,
and only slightly more complex to estimate, as long as density
profile information is known about the dark matter halos under consideration.
Option 3, correcting the halo velocity curve for the effects
of baryonic infall, is, in principle, even more appropriate.
However, the uncertainties associated with this correction are
large.  In the discussion that follows, we will explore
both options 2 and 3, making use of very high-resolution simulation
output that supplies the accurate spatial information
needed for such an analysis.

The Adaptive Refinement Tree (ART) N-body code (\cite{kra97})
reaches high force resolution by refining the
grid in all high-density regions. It allows the identification of
distinct virialized halos as well as halos that
exist as substructure within larger halos.
\cite{kly99} have used the combined results from two
$\Lambda$CDM ($\Omega_0 =1-\Omega_{\Lambda}=0.3$,
$h=0.7$, $\sigma_8=1.0$) ART simulations to explore the
velocity function of halos over a wide range of halo
circular velocities 
(defining $v_c = v_{\rm max}$.) 
The first simulation uses a $60 \hMpc$ box
with a particle mass of $m_{p} = 1.1 \times 10^{9} \hMsun$
and the second simulation uses a $7.5 \hMpc$ box with $m_{p} = 1.7 \times 
10^{7} \hMsun$. They find that the
halo circular velocity function over the range $v_c 
\simeq 20-400$ km s$^{-1}$
is well-described by a power law: 
$\Psi_{\rm halo}(\log v_c) \simeq 0.2 \times 10^{-2.75 \eta_{100}}$,
where $\eta_{100} = \log(v_c/100 \, {\rm km \, s}^{-1})$.
This form of $\Psi_{\rm halo}(\log v_c )$ is shown by the thin solid
line in Figure 2.  We see that near $v_*$, the observations 
are in reasonable agreement with the simulations, although
the density of simulated halos is slightly low.  Correcting this
relation for baryonic infall will help to alleviate this
discrepancy, as we discuss below.

For $v_c \sim$ 400 km s$^{-1}$, the slope of $\Psi_{\rm halo}(\log v_c)$ is shallower than
that observed.  This is not of great 
concern however, since high-velocity halos correspond to 
groups and clusters of galaxies and
should not be compared directly with the observed galaxy velocity functions.  
Modeling galaxies in clusters 
is a difficult problem and
is beyond the scope of this paper.  However, as a first step in 
identifying the appropriate halos for the galaxy velocity function
comparison, we can restrict ourselves to halos which are
``simple'' in the sense that they contain no significant substructure.
Using the $60 \hMpc$ ART simulation and
methods outlined in Bullock et al. (1999a),  we identify halos 
with significant substructure as those containing at
least one 
subhalo 
with $v_c \gsim$ 120 km s$^{-1}$ within the virial radius.
{\footnote{We are unable to push this criterion to a lower value of
$v_c$ due to the incompleteness of our halo catalog, reflecting
the finite resolution of the simulation.}
Our simple-halo velocity function is shown by the 
filled circles connected by the
dashed line in Figure 2.  The errors on these points reflect Poisson
uncertainties.  This first-order correction to the pure halo velocity
function demonstrates a similar fall-off as that observed; however,
the slope remains too shallow for $v_c \gsim$ 400 km s$^{-1}$.  It is likely that
if the substructure criteria for simple halos was more stringent,
for example, if we exclude 
all halos with substructure with $v_c \gsim 70$ km s$^{-1}$, 
then the simple halo velocity function might more closely mirror the
data at high $v_c$.  Higher-resolution simulations would be needed
to check this.

Below about $120$ km s$^{-1}$ the halo number density exceeds the galaxy 
number density.
There are several factors which may contribute to this excess. It is
possible that some fraction of lowest-velocity halos are not associated
with galaxies, as the baryonic material may be ionized and 
unable to cool and form galaxies (\cite{efst92,wei97}).  
Another factor may be selection
effects in luminosity function surveys, in the sense that low surface
brightness galaxies are systematically missed (\cite{spr97,dal98}).
An example of this is LCRS, for which a surface brightness limit was 
imposed upon the spectroscopic sample used to construct the LF (\cite{she96}).
Inclusion of these galaxies can act to steepen the faint end slope of
the observed luminosity function, and hence the observed velocity
function. One intriguing observation is that the orbits of
satellite galaxies exhibit polar anisotropy (\cite{zar97}), and the
suggestion that this may result from the destruction or inhibited
formation of a large population of satellites near the plane of the
disk of spiral galaxies (\cite{zar99}).  Such a scenario could help resolve
the discrepancy, but currently there exists no known physical process
that could accomplish such destruction or inhibition.  

Although a direct comparison with the halo velocity function
is an interesting first step, for a detailed comparison 
one must correct the results of the dissipationless halo
$v_c$ for the effect of baryonic infall.  As a galaxy forms
at the center of a halo, the maximum rotation velocity of
the system increases both due to direct gravitational effects of
the disk and the contraction that infall induces on
the halo.  The overall shift in the velocity 
function will depend on the
nature of the infall and the processes of disk formation; these
are in principle functions of the initial halo $v_c$ and how the
galaxy was assembled including cooling and supernovae feedback.

Assuming that the infall of gas is adiabatic and
that gas infall is halted due to angular momentum support of the 
disk (\cite{fal80}), Blumenthal et al. (1986) 
describes a convenient analytic model for calculating the rotation
curve redistribution during the process of disk formation (see also
Flores et al. 1993 and Dalcanton, Spergel \& Summers 1997).
Mo, Mao, \& White (1997) provide a useful fitting function for the
infall-corrected maximum rotation velocity
of dark matter halos:
\begin{eqnarray}
v_{c} \simeq v_{c}^{\rm halo} 
     \left[ 1 + 4.34m_d - 3.76m_d^2 \right]
          F_V(c_{\rm vir}, \lambda, m_d),
\end{eqnarray}
where $v_{c}^{\rm halo}$ is the maximum rotation velocity of
the halo before infall and
\begin{eqnarray}
F_V(c_{\rm vir}, \lambda, m_d)  = 2.15 \left(\frac{\lambda}{0.1}\right)^
                      {-2.67 m_d - 0.0038/\lambda + 0.2\lambda}
            \left[1 + \frac{\cvir}{17.5} - \left( \frac{\cvir}{54}\right)^2
         - \frac{1.54}{\cvir}\right]\cvir^{-1/2}. 
\end{eqnarray}
Here, $m_d$ is the fraction of the total halo mass that forms the
disk, $\lambda \equiv J|E|^{1/2}G^{-1}M^{-5/2}$ is the dimensionless
angular momentum parameter (where $J$ and $E$ are the total angular
momentum and energy of the halo), and 
$c_{\rm vir} = R_{\rm vir}/R_{\rm s}$ 
describes the nature the dark matter halo density
profile, which is assumed to be of the Navarro, Frenk, \& White (1996) 
form:~\footnote{Although there is disagreement (cf. \cite{kra98,pri98,moo99})  about
the detailed shape of dark matter halo profiles at very
small radii, $r \lsim 0.02$, these very inner regions
are not important for determining $v_{\rm max}$, so the
NFW profile is appropriate for our needs.}
$\rho_{\rm NFW}(r) = \rho_s [(r/R_{\rm s})(1 + r/R_{\rm s})^2]^{-1}.$
(The normalization parameter, $\rho_s$, is 
determined by $v_c^{\rm halo}$ if $R_{\rm s}$ and
$c_{\rm vir}$ are given.)  We have implicitly
assumed that the fraction of the total halo angular momentum in the disk
is equal to the fraction of mass in the disk,  $j_d = m_d$.
Using Equation 10, we may correct the velocity function obtained
from the simulations by using appropriate values for
$\lambda$, $c_{\rm vir}$, and $m_d$.  This may be done halo
by halo, but for simplicity we use the respective averages
of these quantities as a function of $\vch$.

By analyzing the $60 h^{-1}$Mpc ART simulation, Bullock et al. (1999a)
find that the average halo concentration obeys
\begin{equation} 
c_{\rm vir}(v_{c}^{\rm halo})\simeq 13 
\sqrt{\frac{v_{c}^{\rm halo}}{200 \, {\rm km \, s}^{-1}}}
\end{equation}
and the average spin parameter is roughly constant as a function
of the halo circular velocity $<\lambda> \simeq 0.04$ (\cite{bul99b}).
The main uncertainty in this calculation is $m_d$, the fraction
of halo mass that ends up the disk.  This quantity depends on
the details of galaxy formation, including gas cooling and supernovae reheating.
Due to the complexity of the problem, we have used the (fiducial)
semi-analytic models (SAMs) of galaxy formation developed by
Somerville \& Primack (1999) in order to determine a reasonable form
for $m_d(\vch)$.  Using the $\Lambda$CDM cosmology described above,    
we find that the following fitting function
\begin{equation}
m_d(x) \simeq  0.1 \frac{x - 0.25}{1 + x^2},
\label{eqt:md}
\end{equation}
where $x\equiv \vch/(200 \, \rm{km \, s}^{-1})$,
does well in reproducing the average $m_d$ of SAM galaxies
over the range of velocities $v_c \simeq 60-350$ km s$^{-1}$.  
These models assume $\Omega_b = 0.020 h^{-2}$ 
(\cite{bur98}).
Note that $m_d$ rises 
with $v_c$ for $\vch \lsim 200$ km s$^{-1}$;
this is a result of supernova explosions, which act 
to remove gas more effectively
from smaller 
galaxies, 
as proposed in Dekel \& Silk (1986).
After reaching a maximum
of $\sim 0.04$ near $\vch \sim 200$ km s$^{-1}$, the value 
of $m_d$ slowly declines because a smaller fraction of gas
in large halos has time to cool.

Using Equations 10-13 we have corrected the halo velocity
function for the effect of infall.  This correction, which
we will refer to as the SAM infall model, is shown by
the heavy solid curve in Figures 2 and 3.  The curvature in
the SAM infall model is caused by the 
varying behavior of $m_d$ (Eq. 13).  For reference, the medium weight line
shows the result of the infall correction when
the  mass fraction of the disk is held constant 
at $m_d = 0.04$.  The flattening of the SAM-corrected curve at
small velocities is due to the inability
of small disks to retain their gas --- as the disk mass
becomes smaller relative to the mass of the halo, the
correction to the halo circular velocity becomes negligible.
The bend in the SAM-corrected velocity curve at
large $v_c$ is because not all of the gas in large halos
has time to cool.  The SAM infall is truncated at $v_c \gsim 350$ 
km s$^{-1}$ because the
infall calculation is inappropriate for group- and cluster-mass
halos.  We also truncate the curve below $v_c=100$ km s$^{-1}$, 
where $m_d \lsim 0.02$.
Below this value the infall-correction formula (10) 
ceases to be a good fit (\cite{mo98}); however, it is
likely that galaxies with any smaller amount of gas will be of
very low surface brightness and difficult to detect.

It is obvious from this comparison that the halo velocity
function differs markedly from the galaxy velocity function, but
there are avenues of theoretical exploration which may help
in understanding the differences. At
velocities well above $v_*$,  most halos correspond to groups and clusters 
rather than individual galaxies.  Understanding the fall-off
of the velocity function at high $v_c$ will require more detailed 
modeling of galaxy formation within clusters, perhaps
using both semi-analytic and N-body techniques.
Near $v_*$, for the $\Lambda$CDM 
cosmology we explore, the halo 
number density and slope is comparable with the galaxy density to within observational 
errors --- see Figure 4.
The halo density is slightly low without infall correction, and
slightly high with the approximate infall correction we present.
Below $v_*$ the halo density exceeds the galaxy density,
but the tendency for small $v_c$ objects to have small
disk mass fractions due to supernova feedback may help
explain the discrepancy: because they have low luminosity and low surface
brightness, many low-$v_c$ galaxies will be missed in the luminosity
functions we started from. If, however, the discrepancy at the low-$v_c$ end 
is not purely due to selection effects, it may turn out to pose a real
challenge for theory.
More detailed modeling of 
redshift survey selection effects and of small-velocity galaxies,
including consistent treatments of gas cooling, baryonic infall,
supernova feedback, and disk surface brightnesses will be needed
to explore this problem in detail.

\section{Discussion and Conclusions}
\label{sec:conc}

A main goal of this work was to evaluate the robustness with which
$\Psi( \log v_c)$ can currently be estimated.
While morphological
information was incorporated in converting the SSRS2 luminosity
function to a velocity function, a key result of this detailed
analysis is that treating the entire population as spirals does not
significantly alter the resulting velocity function.\footnote{Unless
there is a large error in the FJ zeropoint, in which case it may be
necessary to treat E's separately in order to accurately reproduce the
high velocity end.}  
Furthermore, 
while the normalization of the velocity
function remains poorly constrained ($\sim$30\% 
variance in $\Psi_*$
among surveys (excluding $K$), and a factor of 2 variance in 
$\Psi(\log v_c)$ at $v_c = 240$ km s$^{-1}$),
the {\it shape} of the velocity function is similar
for all input luminosity functions. Both the shape and normalization are
also consistent within the errors with the velocity function derived by Shimasaku (1993).

The key benefit of our approach is that the models needed to connect
N-body simulations and observations become much less complex when we use 
observed TF relations and extinction corrections instead of trying to reproduce 
these functions via the semi-analytic models. This contrasts with SAMs, 
which output modelled $l-v$ relations that are dependent on tunable
model parameters (e.g., star formation timescales, supernovae feedback, etc.)
for comparison with observations. Another benefit is that, by converting
to velocity, we provide a single target function for the models to 
attempt to reproduce --- in contrast with luminosity functions in different bands
from various redshift surveys. Hopefully, this will be of value in simplifying comparison 
with simulations from different groups.

The main sources of uncertainty limiting the precision with which the
velocity function can be constructed via this approach are:

\begin{enumerate}
  \item{Large scatter in the reported values of $\Phi_*$, possibly due
    to local deviation from mean density.}
 
  \item{Selection bias at the faint end of the luminosity function,
    which may act to flatten the faint end slope.}

  \item{The limited velocity range over which the TF relation is well-calibrated.}

  \item{Uncertainty in the TF relation. Beyond the statistical errors, there
    are indications of several potentially significant biases in current 
    relations. In particular, Willick and Courteau (private communication) find
    that using the circular velocity at two disk scale lengths (as determined
    by disk+bulge fitting) reduces scatter in the TF relation and also can significantly
    alter the slope relative to other methods of determining $v_c$.  }

  \item Uncertainty regarding the extinction correction, and also over-simplified treatment 
	of extinction by averaging over inclination.

  \item{Uncertainty in the zeropoint of the FJ relation.  A
    change in the zeropoint could noticeably alter the velocity
    function above $v_*$.}

  \item{Lack of a detailed understanding of the correspondence between
    $\sigma$ in ellipticals and $v_c$ in spirals. While $v_c=\sqrt2
    \sigma$ may be true on average for bright ellipticals, scatter
    in this relation can also alter the velocity function above
    $v_*$.}

  \item Intrinsic scatter in the TF and FJ relations.

\end{enumerate}

Use of next-generation surveys such as Anglo-Australian 2-degree
Field (2dF) and the Sloan Digital Sky
Survey (SDSS) will reduce the cosmic variance of $\Phi_*$.  Further, an 
alternative approach to this technique would be the direct
construction of the velocity function from a galaxy survey designed to obtain
both photometry and slit-based spectral linewidths. This could be achieved by
a slitmask survey of a volume-limited subset of the SDSS or 2dF 
samples.\footnote{While 2dF
and SDSS both obtain spectral information, the fiber-fed spectrographs
only collect information on galactic centers, and hence do not provide 
information on the rotation curves of spiral galaxies.} 
Bypassing the $l-v$ relations would significantly reduce
uncertainty in the derived velocity function, although it would be
necessary to separate the spiral and elliptical populations so as to
treat rotationally- and thermally-supported systems correctly.
Finally, with the multi-color photometry of SDSS, it will be
possible to measure inclinations and 
correct for extinction effects on a galaxy-by-galaxy basis prior to 
construction of the luminosity function.

 In the meantime, we have demonstrated that existing data is
sufficient to construct 
the velocity function accurate to within roughly a factor of two at
a given velocity, which is suitably accurate for comparision with 
predictions from cosmological N-body simulations. By comparison with
one such $\Lambda$CDM simulation, we have illustrated the usefulness of
this approach. The main sources of uncertainty limiting 
the precision with which the velocity function can be estimated theoretically are:

\begin{enumerate}

\item The degree to which halos with very large and very small $v_c$ 
should be associated with galaxies.

\item Uncertainties associated with correcting the $v_c$ of measured 
halos for the effect of baryonic infall.

\end{enumerate}

There is reasonable agreement between the observations and simulations 
near $v_*$, and the exploration of these uncertainties may help explain 
the large excess of systems in the simulations below  
$\sim$120 km s$^{-1}$ and the slope of the velocity function above $v_*$.  
This poses an interesting challenge
for models of galaxy formation to address.  A key test will be to see
if the incorporation of baryonic infall and cooling physics leads to a
theoretical galaxy velocity function consistent with 
those observed.

\acknowledgements
\section{Acknowledgements}
	The authors would like to thank the anonymous referee for suggestions
which significantly improved this paper, especially regarding the need to include
passband corrections.  We also thank Dennis Zaritsky, Raja
Guhathakurta, Sandra Faber, Stephane Courteau, Jeff Willick, and Chris
Kochanek for helpful discussions regarding this research. AHG and KAW
acknowledge support from the NSF Graduate Research Fellowship Program.
JSB, TSK, and JRP were supported by a NASA ATP grant.

\appendix

\section{Appendix}
Starting with the Schechter function,
\begin{equation}
 \Phi(L) dL = \Phi_* \left(\frac{L}{L_*}\right)^\alpha
 \exp{\left(-\frac{L}{L_*}\right)} d\left(\frac{L}{L_*}\right),
\end{equation}
and the relation 
\begin{equation}
L=A x^n,
\end{equation}
where $A$ is a constant, we define 
\begin{equation}
\tilde{\Psi}(x) dx=\Phi(L) dL \hspace{4.5cm}
\end{equation}
\begin{equation}
\hspace{3cm}   = \Phi_* \left(\frac{x}{x_*}\right)^{n \alpha}
 \exp{\left[-\left(\frac{x}{x_*}\right)^n\right]} n
 \left(\frac{x}{x_*}\right)^{n-1} d\left(\frac{x}{x_*}\right).
\end{equation}
\begin{equation}
\hspace{2cm}   = n \Phi_* \left(\frac{x}{x_*}\right)^{n \alpha +n -1}
 \exp{\left[-\left(\frac{x}{x_*}\right)^n\right]} d\left(\frac{x}{x_*}\right).
\end{equation}
Thus,
\begin{equation}
 \tilde{\Psi}(x) dx = \tilde{\Psi}_* \left(\frac{x}{x_*}\right)^{\beta}
 \exp{\left[-\left(\frac{x}{x_*}\right)^n\right]} d\left(\frac{x}{x_*}\right),
 \hspace{0.5cm}
\end{equation}
where $\beta\equiv n(\alpha+1)-1$ and $\tilde{\Psi}_*\equiv n\Phi_*$.

Inclusion of a magnitude dependent extinction correction of the 
form,
\begin{equation}
M^{cor} = \gamma^{-1} (M + C),
\end{equation}
where $\gamma$ and $C$ are constants, is equivalent to modifying the
power and zeropoint of the $L$-$x$ relation. Specifically, for the
case of a TF relation (which in general will be extinction corrected)
of the form,
\begin{equation}
M^{cor}=a - b (\log 2 v_c - 2.5 ),
\end{equation}
in order to relate $v$ to the observed magnitude,$M$, from a
luminosity function survey (which in general will {\it not} be
extinction corrected), we substitute
\begin{equation}
M^{cor} = \gamma^{-1} (M + C) = a - b (\log 2 v_c - 2.5),
\end{equation}
or,
\begin{equation}
M = (\gamma a - C)  -  \gamma b (\log 2 v_c - 2.5)
= a^{'} -b^{'} (\log(\Delta v) - 2.5).
\end{equation}
The resulting modified TF relation is thus of the same form as the
original, but with a new slope and offset that are related to the old
ones by
\begin{equation}
a^{'} = \gamma a - C,
\end{equation}
and
\begin{equation}
b^{'} = \gamma b.
\end{equation}

\clearpage
\begin{deluxetable}{ccrcllrll}
\footnotesize
\tablecaption{A sample of published field luminosity functions. \label{tbl-1}}
\tablewidth{0pt}
\tablehead{
\colhead{Survey} &\colhead{Band} & \colhead{\# Galaxies} & \colhead{Type} &
\hfil M$_*$\hfil &  \hfil$\alpha$\hfil & \hfil $\Phi_*$ ($h^3$ Mpc$^{-3}$)\hfil & M$_{low}$ & M$_{high}$
}

\startdata
SSRS2          & $B_{ssrs2}$ & 5306  & All    & -19.43$\pm$0.06 & -1.12$\pm$0.05 & 12.8$\pm$2.0$\times10^{-3}$  & -21.1  & -15.58 \nl
               &             &       & E/S0   & -19.37$\pm$0.11 & -1.00$\pm$0.09 & 4.4$\pm$0.8$\times10^{-3}$   & -21.1  & -14.   \nl
               &             &       & Sp     & -19.43$\pm$0.08 & -1.11$\pm$0.07 & 8.0$\pm$1.4$\times10^{-3}$   & -21.1  & -15.58 \nl
               &             &       & Irr    & -19.78$\pm$0.50 & -1.81$\pm$0.24 & 0.20$\pm$0.08$\times10^{-3}$ & -21.1  & -14.   \nl
APM            & $b_J$       & 1769  & All    & -19.50$\pm$0.13 & -0.97$\pm$0.15 & 14.0$\pm$1.7$\times10^{-3}$  & -21.25 & -15.5  \nl
UKST           & $b_J$       & 2500  & All    & -19.68$\pm$0.10 & -1.04$\pm$0.08 & 17.0$\pm$3.0$\times10^{-3}$  & -21.5  & -14.5  \nl
CfA            & $Z$         & 9063  & All    & -18.90          & -1.02          & 20.1$\pm$5.0$\times10^{-3}$  &        &        \nl
LCRS           & $r_g$/$R_C$ & 22743 & All    & -20.29$\pm$0.02 & -0.70$\pm$0.05 & 19.0$\pm$1.0$\times10^{-3}$  & -22    & -18.   \nl
               &             &       & Clan 1 & -20.29$\pm$0.07 &  0.51$\pm$0.14 & 0.40$\pm$0.02$\times10^{-3}$ & -22.5  & -18.   \nl
               &             &       & Clan 2 & -20.23$\pm$0.03 & -0.12$\pm$0.05 & 6.9$\pm$0.5$\times10^{-3}$   & -22.5  & -16.5  \nl
               &             &       & Clan 3 & -19.89$\pm$0.04 & -0.31$\pm$0.07 & 8.5$\pm$0.1$\times10^{-3}$   & -22.5  & -16.5  \nl
               &             &       & Clan 4 & -19.86$\pm$0.05 & -0.65$\pm$0.08 & 7.3$\pm$0.2$\times10^{-3}$   & -22    & -16.5  \nl
               &             &       & Clan 5 & -19.95$\pm$0.09 & -1.23$\pm$0.10 & 1.9$\pm$0.6$\times10^{-3}$   & -21.5  & -16.5  \nl
               &             &       & Clan 6 & -20.10$\pm$0.16 & -1.93$\pm$0.13 & 0.7$\pm$0.5$\times10^{-3}$   & -21    & -17.   \nl
Gardner et al. & $K$         & 567 & All & -23.30$\pm$0.30 & -1.00$\pm$0.20 & 14.4$\times10^{-3}            $   & -25    & -20.5  \nl
\enddata
\tablecomments{For LCRS, the survey data was obtained using $r_g$, but calibrated to $R_C$. M$_{low}$ and M$_{high}$ denote
the magnitude range over which the given Schechter function is a good fit to the data.}
\end{deluxetable}

\clearpage
\begin{deluxetable}{lcllrr}
\tablecaption{Tully-Fisher parameters. \label{tbl-2}}
\tablewidth{0pt}
\tablehead{
\colhead{} &\colhead{Band} & \colhead{a} & \colhead{b} & $v_{low}$ & $v_{high}$ 
}

\startdata 
Yasuda et al. 1997       & $B_T$ & $-18.71\pm0.11$ & $6.76\pm0.63$ & 67 & 276 \nl 
Richter et al. 1987      & $B_T$ & $-18.45\pm0.39$ & $7.17\pm0.20$ & $\sim$10 & $\sim$280 \nl 
Hoffman et al. 1996 (Irr)& $B_T$ & $-18.13\pm0.70$ & $6.50\pm0.63$ & $\sim$10 & $\sim$80 \nl
Courteau 1997            & $r_s$/$r_g$ & $-20.00\pm0.03$ & $6.17\pm0.28$ & $\sim$90 & $\sim$350 \nl 
Malhotra et al. 1996     & $K$ & $-21.41\pm0.11$ & $8.59\pm0.67$ & 117 & 273 \nl
de Grijs \& Peletier 1999 & $K$ & $-22.48\pm1.58$ & $8.09\pm0.52$ & $\sim$90  & $\sim$310 \nl
Tully et al. 1998	 & $K'$ & $-22.67$	&   8.73	&  $\sim$50	& $\sim$280 \nl

\tablecomments{The listed values are coefficients to the equation
M=a-b ($\log(\Delta v$)-2.5), and have been normalized to $H_0$=100 km
s$^{-1}$. The value listed for Courteau 1997 corresponds to the
determination using $v_c$ at 2.2 optical scale lengths for the
Courteau-Faber `quiet Hubble flow' sample. The imaging for Courteau 1997 was obtained with
a Spinrad $r$ filter, $r_s$, but calibrated to $r_g$. In the last two columns $v_{low}$ and
$v_{high}$ indicate the limits of the velocity range spanned by the data used to construct
these TF relations.}

\enddata
\end{deluxetable}
\clearpage

\begin{deluxetable}{rlll}
\tablecaption{Passband Conversions. \label{tbl-3}}
\tablewidth{0pt}
\tablehead{
\multicolumn{3}{c}{Transformation} & \colhead{Reference}}

\startdata 
$B_{SSRS2}\!\!\!$ & $=$ $B_T + 0.26$ & \hfil & Alonso et al. 1993
\nl
$b_J\!\!\!$ & $\approx$ $B_T  + 0.06$ & \hfil   & Cole et al. 1994; Alonso
et al. 1993   \nl
$r_{LCRS}\!\!\!$ & $\approx$ $r_{Courteau} + 0.33$& \hfil   & Djorgovski
1985; Fukugita et al. 1995 \nl
$r_{LCRS}\!\!\!$ & $\approx$ $R_C  + 0.36$ & \hfil  & Fukugita et al. 1995
\nl

\tablecomments{The color transformations used to place $l-v$ relations,
luminosity 
functions, and extinction corrections on the same filter system. } 
\enddata
\end{deluxetable}
\clearpage

\begin{deluxetable}{llllr}
\tablecaption{Velocity Function Parameters \label{tbl-4}}
\tablewidth{0pt}
\tablehead{
\colhead{Survey} & \colhead{TF Relation} & \colhead{v$_*$ (km s$^{-1}$)} &
\colhead{$\beta$} &\colhead{$\Psi_*$ (Mpc$^{-3} h^{3}$)}
}

\startdata

SSRS2      & Yasuda   & $247\pm 7$ & $-1.30\pm0.13$ & $7.3 \pm1.4 \times10^{-2}$ \nl
SSRS2      & Richter  & $261\pm15$ & $-1.32\pm0.13$ & $7.8 \pm1.2 \times10^{-2}$ \nl
APM        & Yasuda   & $253\pm 8$ & $-0.93\pm0.37$ & $8.0 \pm1.3 \times10^{-2}$ \nl
UKST       & Yasuda   & $271\pm 9$ & $-1.10\pm0.20$ & $9.7 \pm2.0 \times10^{-2}$ \nl
LCRS       & Courteau & $215\pm 2$ & $-0.30\pm0.12$ & $10.3\pm0.7 \times10^{-2}$ \nl
Gardner    & Malhotra & $265\pm11$ & $-1.00\pm0.68$ & $11.3\pm4.8 \times10^{-2}$ \nl

\tablecomments{The listed values are the derived parameters for the
velocity function using an assortment of surveys and Tully-Fisher
relations.  
For LCRS, the published single Schechter function fit is
used (\cite{lin96}), whereas in Figure 2 the fits to the individual
clans are used. Also, no error is given for $\Phi_*$ in the Gardner
survey; {\it the error given here is an estimate
based upon the relative number of galaxies in the Gardner sample
compared to other surveys}. Even with no error in $\Phi_*$ however,
the error in $\Psi_*$ resulting from statistical uncertainty in the 
$K$-band TF relation would be $8.7\times10^{-3}$ Mpc$^{-3} h^{3}$.
Note that $\Psi_*$ corresponds to
the ($\log v_c$) velocity function, $\Psi(\log v_c)$, the related
quantity for $\tilde{\Psi}(v_c)$ is $\tilde{\Psi}_* = \Psi_* / \ln(10) $ (see Eq. 3).
}

\enddata
\end{deluxetable}
\clearpage

\clearpage
\begin{figure}
\plotone{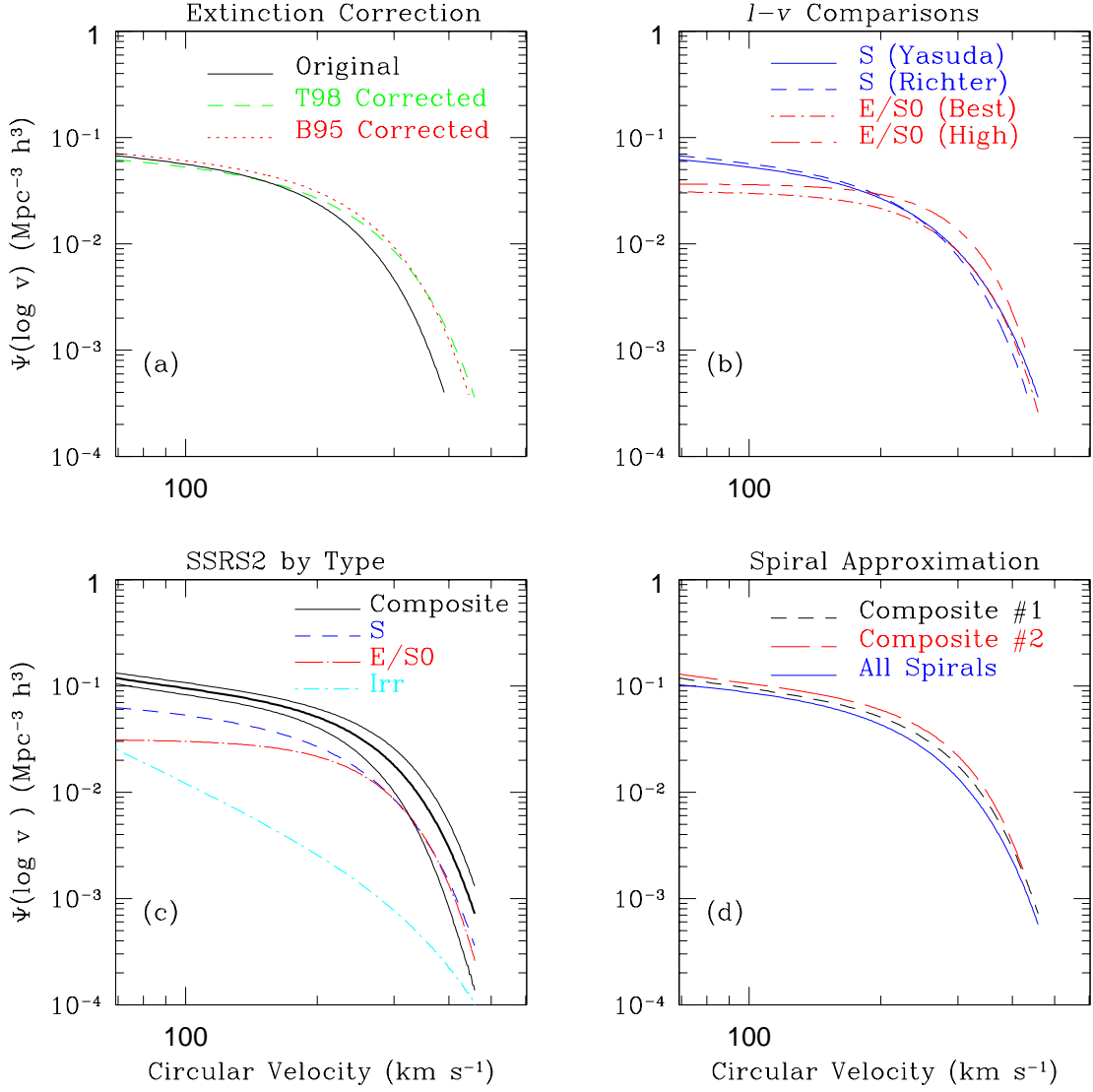}
\figcaption[fig1.new.eps]{
Various SSRS2 velocity function estimates: 
(a) Effect on spiral velocity function of correction for
internal extinction, using both T98 and B95 extinction laws. The net 
effect of both is to shift the velocity function up by $\sim$30 km s$^{-1}$. 
(b) Different choices of $l-v$ relation parameters (see text).
(c) Composite velocity function and
contributions by each morphological class. The Yasuda  TF relation and
``Best'' FJ relation are used.  One standard deviation errors are represented by the
two thin lines.
(d) Differences in the derived velocity function under the assumption that
all galaxies are spirals (short dashed line).  Composite \#1 (solid line) is
the velocity function generated using the Yasuda, Fukugita, \& Okamura (1997)
 TF relation and the "Best'' FJ relation. For Composite \#2 (long dashed line), 
the "High'' FJ relation is used instead to illustrate the effect of uncertainty in the 
FJ relation. 
\label{fig1}}
\end{figure}
\clearpage
\begin{figure}
\plotone{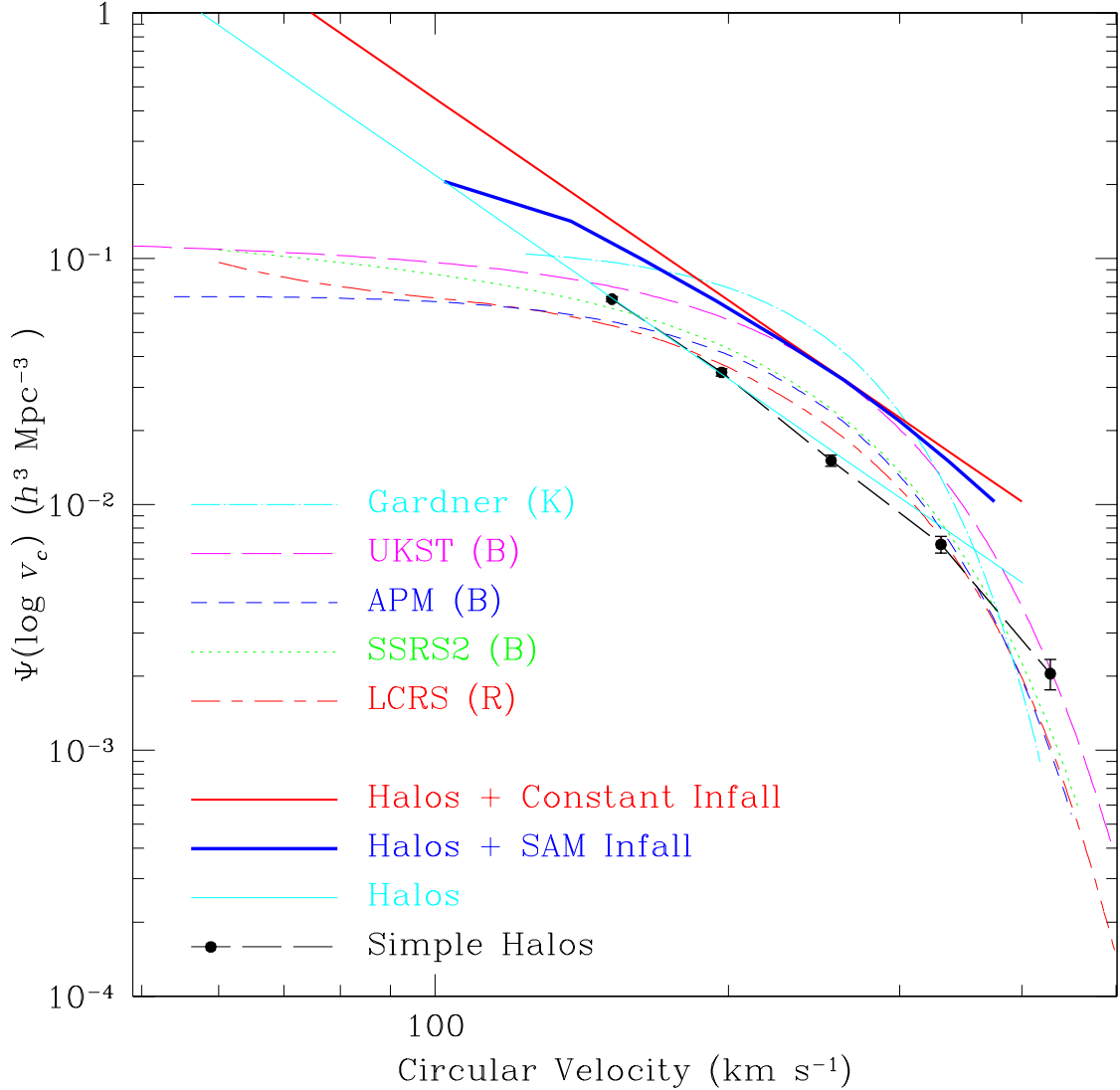}
\figcaption[fig2.new.eps]{Velocity functions for five different surveys
in three different bands generated using the spiral approximation. Each is
plotted over the range for which a Schechter luminosity function is a good
fit to the LF data.  The thin solid line reflects the dark matter
halo velocity functions as determined from N-body simulations.
The dashed line connecting the points 
is the corresponding ``simple halo'' 
velocity function, which neglects
all halos in the simulation with signifcant substructure. 
Also shown is the halo velocity function
corrected for the effect of baryonic infall, using a constant
disk mass fraction approximation 
(straight thin line) 
and a more complicated
assumption based on results from semi-analytic models of
galaxy formation (SAMs, bold line). \label{fig2}}
\end{figure}
\clearpage
\begin{figure}
\plotone{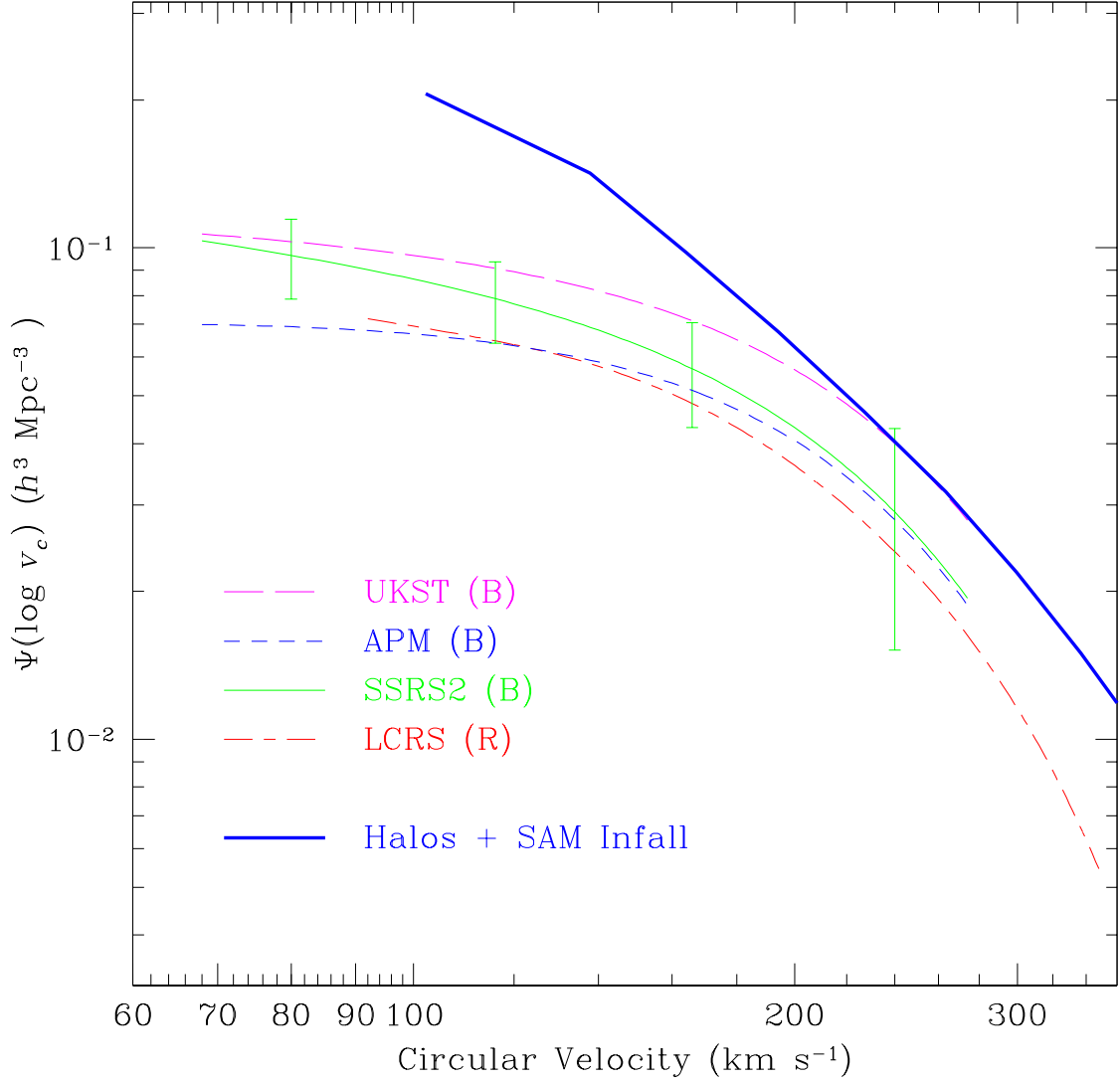}
\figcaption[fig3.new.eps]{ Same as Figure 2, except that 
the observed velocity functions are now only plotted in the region in which both 
the luminosity function and TF relation used to generate each velocity function 
are well-constrained by the data, and the $K$-band data is no longer plotted. For SSRS2, 
the line style has been changed to solid and error bars have been added. Also, for clarity 
only the SAM infall model is plotted.
\label{fig3}}
\end{figure}
\clearpage
\begin{figure}
\plotone{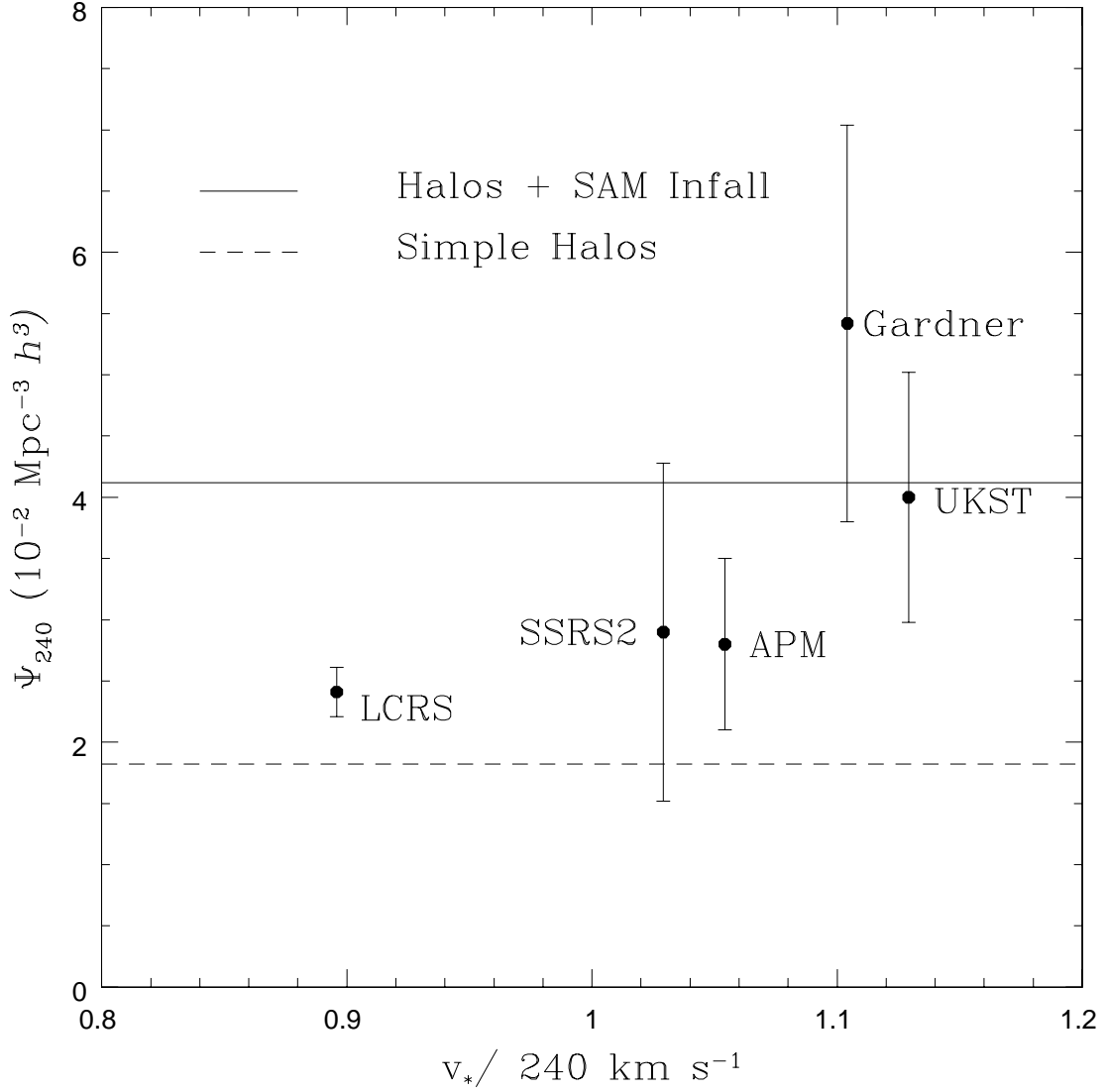}
\figcaption[fig4v.new.eps]{Comparison of 
$\Psi_{240}=\Psi(\log v_c)$ at $v_c=240$ km s$^{-1}$ 
for five surveys. The error bars are the formal statistical errors and do
not include likely systematic effects.
The x-axis is given in units of each survey's $v_*$.
\label{fig4}}
\end{figure}

\end{document}